\documentclass[11pt]{article}
\usepackage{amssymb}
\usepackage{amsmath}
\usepackage{amstext}
\usepackage{graphicx,epsfig}
\usepackage{epsfig}
\usepackage{verbatim} 
\usepackage{fancybox}
\usepackage{color}
\usepackage{ulem}
\usepackage{enumitem}
\usepackage{subfigure}
\usepackage{bbm}
\usepackage{parskip}
\usepackage{dsfont}
\usepackage[numbers,sort&compress]{natbib}

\newcommand{\Comment}[1]{{}}
\definecolor{darkblue}{rgb}{0.15,0.35,0.55}
\definecolor{reddish}{rgb}{0.65, 0.2, 0.2}
\usepackage[linktocpage=true]{hyperref}
\hypersetup{
colorlinks=true,
citecolor=darkblue,
linkcolor=reddish,
urlcolor=darkblue,
pdfauthor={},
pdftitle={},
pdfsubject={}
}

\newcommand{\be}{\begin{equation}}
\newcommand{\ee}{\end{equation}}
\newcommand{\bea}{\begin{eqnarray}}
\newcommand{\eea}{\end{eqnarray}}
\newcommand{\beas}{\begin{eqnarray*}}
\newcommand{\eeas}{\end{eqnarray*}}
\newcommand{\nn}{\nonumber \\ }

\def\({\left(}
\def\){\right)}

\newcommand{\rd}{{\rm d}}

\def\gsim{ \lower .75ex \hbox{$\sim$} \llap{\raise .27ex \hbox{$>$}} }
\def\lsim{ \lower .75ex \hbox{$\sim$} \llap{\raise .27ex \hbox{$<$}} }

\title{}
\author{}

\numberwithin{equation}{section}

\begin{document}
%
~
\vspace{2truecm}
\begin{center}
{\LARGE \bf{Spontaneously Broken Gauge Theories and the Coset Construction}}
\end{center} 

\vspace{1.3truecm}
\thispagestyle{empty}
\centerline{\Large Garrett Goon,${}^{\rm a}$ Austin Joyce,${}^{\rm b}$ and Mark Trodden${}^{\rm a}$}
\vspace{.7cm}

\centerline{\it$^{\rm a}$Center for Particle Cosmology, Department of Physics and Astronomy,}
\centerline{\it University of Pennsylvania, Philadelphia, PA 19104, USA}

\vspace{.3cm}

\centerline{\it ${}^{\rm b}$Enrico Fermi Institute and Kavli Institute for Cosmological Physics}
\centerline{\it University of Chicago, Chicago, IL 60637}
\vspace{.4cm}

\begin{abstract}
\vspace{.03cm}
\noindent
The methods of non-linear realizations have proven to be powerful in studying the low energy physics resulting from spontaneously broken internal and spacetime symmetries.  In this paper, we reconsider how these techniques may be applied to the case of spontaneously broken gauge theories, concentrating on Yang--Mills theories. We find that coset methods faithfully reproduce the description of low energy physics in terms of massive gauge bosons and discover that the St\"uckelberg replacement commonly employed when treating massive gauge theories arises in a natural manner. Uses of the methods are considered in various contexts, including generalizations to $p$-form gauge fields. We briefly discuss potential applications of the techniques to theories of massive gravity and their possible interpretation as a Higgs phase of general relativity.
\end{abstract}

\newpage

\tableofcontents

\section{Introduction}  

Gauge symmetries form a pillar of modern physics and as such they have been studied and interpreted in myriad ways.  Here we focus on a treatment of gauge fields in the context of non-linear realizations.  The basis for this approach is the observation that the global part of a gauge symmetry acts linearly on gauge fields, while the local symmetries act non-linearly.  That is, the transformation for a Yang--Mills (YM) one-form gauge field, $A$, schematically given by $A\mapsto U(A+\rd) U^{-1}$, is generally non-linear due to the presence of the second term, but becomes linear in the global limit where $U$ is independent of $x$. This allows us to think of the gauge field as a Goldstone field non-linearly realizing the gauge symmetry and to apply standard coset construction techniques to build its action.\footnote{It is quite surprising that this is possible; after all gauge symmetry is merely a redundancy of description and not a physical symmetry. Nevertheless, we will see that decomposing the gauge symmetry into an infinite number of global transformations will allow us to construct gauge fields as Goldstones.}

The techniques of non-linear realizations were developed in the context of spontaneously broken global symmetries~\cite{Callan:1969sn,Coleman:1969sm, volkov}.  Given a spontaneous symmetry breaking (SSB) pattern in which a global symmetry group $G$ is broken to a preserved subgroup $H$, denoted $G\to H$, these methods generate all terms which can appear in the low energy effective action used to describe the physics of the broken phase \cite{D'Hoker:1994ti}.  The preserved symmetries of $H$ act linearly on the resulting low energy degrees of freedom, while those in $G/H$ act non-linearly. The coset construction has recently seen a resurgence of interest (see~\cite{Watanabe:2012hr,Hinterbichler:2012mv,Goon:2012dy,Creminelli:2013fxa,Nicolis:2013sga,Nicolis:2013lma,Endlich:2013vfa,Watanabe:2014fva,Kampf:2014rka,Hinterbichler:2014cwa} for a variety of novel applications).

The power of coset methods lies in their generality.  For example, in the case of (internal) global symmetry breaking one only needs the knowledge of the breaking pattern to derive the universal form of the low energy, infrared (IR) action. Long-wavelength features are insensitive to the detailed high energy microphysics responsible for breaking the symmetries.  The goal of this paper is to elucidate the analogous result for gauge theories.  That is, to determine the gross features of the Higgs phase of a gauge theory only from the knowledge of the breaking pattern, while remaining agnostic about the theory's ultraviolet (UV) completion.

Gauge fields are not typically interpreted as low energy degrees of freedom arising from spontaneous symmetry breaking, but the coset techniques described above are still useful in this context.  A typical gauge group, which we denote $G_{\rm local}$, is infinite dimensional and includes global transformations as a subgroup, denoted $G_{\rm global}$.  As noted previously, the action of $G_{\rm global}$ on gauge fields is linear while the action of an element in $G_{\rm local}/G_{\rm global}$ is non-linear.  Applying the coset methods to the ``breaking" pattern $SU(N)_{\rm local}\to SU(N)_{\rm global}$, for example, one can generate the YM action \cite{Ivanov:1976pg,Borisov:1974bn, Kosinski:1976pd}. The Einstein--Hilbert  action can be derived similarly by considering the diffeomorphism group~\cite{Ogievetsky:1973ik}.

Non-linear realization techniques have previously been used to study Higgs phases of gauge theories, but this was done by first applying coset methods to global symmetry breaking patterns and then gauging the resulting theory of Goldstone bosons by hand \cite{Burgess:1992gx,Weinberg:1996kr}.  Here, we instead systematically construct the appropriate actions for broken gauge theories entirely in the coset framework.  This alternate route has gauge bosons built in from the start and makes contact with a greater number of conceptual and technical aspects of coset methods, such as the subtleties of spacetime symmetries due to ``Inverse Higgs" effects, and the use of cohomological methods in finding Wess--Zumino terms.

Before studying the broken phase of gauge theories, we review the standard coset construction for both internal and spacetime symmetry groups and then perform the construction of unbroken Yang--Mills theory in this language. Along the way, we provide interesting demonstrations of how various facets of gauge theories are expressed in the coset language. For instance, the search for Wess--Zumino terms leads us to the construction of Chern--Simons terms.  

We then turn to the main case of interest and demonstrate that these techniques are also applicable to Yang--Mills theories which truly exhibit spontaneous symmetry breaking.  That is, given a SSB pattern in which $G_{\rm global}$ is at least partly broken, coset methods correctly reproduce the fact that the low energy degrees of freedom are a mix of massive and massless gauge bosons, where the precise mixture depends on the breaking pattern.  Further, we find that coset techniques automatically employ the St\"uckelberg trick commonly used for treating massive gauge bosons.

In the final section we discuss applications and generalizations of these methods.  We consider scenarios in which different residual symmetries are preserved after SSB and work out one such case in detail, confirming that we accurately reproduce known results. We also discuss the generalization to the case of $p$-form gauge fields where the St\"uckelberg realization of the $p$-form gauge symmetry arises naturally. Finally, we discuss the potential use of coset methods in treating anomalous gauge theories.

There are many generalizations and applications of the formalism we present. One such application is to the spontaneous breaking of gauge symmetries in the non-relativistic setting. Although throughout we assume that Poincar\'e symmetry is preserved, our results generalize straightforwardly to the non-relativistic arena. Another possible application of the coset formalism is to investigate the Abelian vector duality presented in~\cite{deRham:2014lqa}, which we discuss. Finally, we briefly discuss the potential application of our methods to theories of gravity. In this context, coset methods can be used to study generic IR properties of gravitational Higgs mechanisms in a systematic way which is insensitive to the precise UV mechanism.    In particular, we are interested in studying breaking patterns which can give rise to ghost-free de Rham--Gabadadze--Tolley (dRGT) \cite{deRham:2010kj} massive gravity and exploring whether the dRGT interactions are special in some way when compared to the generic terms one generates.  We explore this possibility in detail in a companion paper~\cite{GHJTGravitypaper}.

\section{General Coset Methods}\label{Sec:General Coset Methods}

In this section, we review the machinery of non-linear realizations for treating the spontaneous breaking of both internal and spacetime symmetry groups.  The procedure for the internal case was originally developed in~\cite{Callan:1969sn,Coleman:1969sm} and is discussed nicely in~\cite{Zumino:1970tu,Weinberg:1968de,D'Hoker:1994ti,Ogievetsky:1974}. The construction was generalized to broken spacetime symmetries in~\cite{volkov} and techniques and subtleties relevant to the spacetime case can be found in~\cite{Ogievetsky:1974,Ivanov:1975zq,Low:2001bw,McArthur:2010zm,Nicolis:2013sga,Brauner:2014aha}.  Both cases are reviewed in~\cite{Goon:2012dy,Hinterbichler:2012mv}, which we follow.

\subsection{Internal Symmetry Breaking}
\label{internalsymms}
Consider the spontaneous breaking of an internal symmetry group, $G$, to a subgroup $H$. Let $V_{I}$, $I\in\{1,\ldots,\dim H\}$, be the generators of the preserved subgroup, $H$, and all other generators be denoted by $Z_{a}$, $a\in\{1,\ldots,\dim G/H\}$. We refer to $V_{I}$'s and $Z_{a}$'s as the ``unbroken" and ``broken" generators, respectively. Representative coset elements, $\tilde{g}\in G/H$, are written in the canonical form $\tilde{g}=\exp(\xi^{a}Z_{a})$. From Goldstone's theorem, there are as many Goldstone bosons as there are broken generators and we identify the associated fields with the coordinates of the coset space $G/H$.  The coset elements are  maps from spacetime, denoted $\mathcal{M}$, to the coset space, $\tilde{g}(x):\mathcal{M}\to G/H$, and the Goldstone fields are the $\xi^{a}(x)$'s.
 
 Every group element $g\in G$ defines a symmetry transformation of the fields, $g:\xi^{a}\to \xi'^{a}$, defined through
 \begin{align}
 g\exp(\xi^{a}(x)Z_{a})&=\exp(\xi'^{a}(x)Z_{a})h(g,\xi^{a}(x))\label{Goldstonetransformation}\ ,
 \end{align}
 where $h(g,\xi^{a}(x))$ is an element of $H$.  Generally, the transformation $\xi^{a}\to\xi'^{a} $ is complicated and non-linear, but in the limit that $g\subset H$ the relationship becomes linear.\footnote{Strictly, speaking this is only if commutators between broken and  unbroken generators never contain any unbroken generators.  The algebras we consider obey this restriction.}
 
In order to build actions for the $\xi^{a}$ fields we employ the Lie algebra-valued Maurer--Cartan form 
\begin{equation*}
\tilde{g}^{-1}\rd \tilde{g}\equiv \Omega=\Omega_{Z}+\Omega_{V}=\Omega_{Z}^{a}Z_{a}+\Omega_{V}^{I}V_{I} \ .
\end{equation*}
The utility of $\Omega$ is that it transforms nicely under \eqref{Goldstonetransformation}, where $\Omega_{Z}$ transforms homogeneously and $\Omega_{V}$ transforms as a connection,
 \begin{align}
 g:\begin{cases}\Omega_Z&\longmapsto h(x) \,\Omega_Z \,h^{-1}(x)\\ \Omega_V &\longmapsto h(x)\,(\Omega_V+\rd)\,h^{-1}(x)\end{cases}\ .\label{maurercartantransformation}
 \end{align}
 
For a $d$-dimensional spacetime, one then builds a $G$-invariant $d$-form lagrangian for the Goldstone fields by combining together factors of $\Omega_{Z}^{a}$'s using the exterior product\footnote{Equivalently, the forms $\Omega_{Z}^{a}$ can be used to construct a covariant derivative for the Goldstone fields: $\rd x^\mu{\cal D_\mu}\xi^a = \Omega_Z^a$, which transforms covariantly.} and contracting indices appropriately so that the result is $H$-invariant, in the sense that the final $d$-form is invariant under  \eqref{maurercartantransformation}.   Other matter fields couple to the Goldstones via the covariant derivative whose connection is defined by $\Omega_{V}$.

The above procedure of creating $d$-form lagrangians only constructs terms {\it strictly} invariant under the relevant symmetries and therefore may miss certain ``Wess--Zumino" (WZ) terms which shift by a total derivative under the symmetries. These terms can also appear in the action, but require a higher dimensional construction \cite{Witten:1983tw, D'Hoker:1994ti}.   Finding a WZ term is equivalent to a cohomology calculation: one looks for an exact, $H$-invariant $(d+1)$-form\footnote{One imagines that the form $\alpha$ is constructed either on a $(d+1)$-dimensional spacetime or on $G$ itself so that the $(d+1)$-form is well defined and not simply automatically zero.} $\alpha=\rd\beta$, built out of the $\Omega_Z^a$, such that $\beta$ is \textit{not} itself $H$-invariant.  Instead, $\beta$ shifts by a closed form under the symmetries, so that $\alpha=\rd\beta$ is still strictly invariant, and thus $\beta$ represents a perfectly fine term that can be added to the $d$-dimensional action. That $\beta$ is not strictly invariant is equivalent to the statement that it is not built out of the $\Omega_Z^a$ building blocks and is therefore a term that would be missed had we restricted ourselves to only searching for $d$-forms.  (See  \cite{D'Hoker:1994ti,D'Hoker:1995it,deAzcarraga:1997gn,deAzcarraga:1998uy} for more on the cohomological aspects of WZ terms.)
 
This higher dimensional construction can help elucidate the quantum mechanical properties of Wess--Zumino terms~\cite{Witten:1983tw}.  Given the appropriate $(d+1)$-form $\alpha$, one can compactify spacetime so that it encloses a $(d+1)$-dimensional ball ${\cal B}$, with $\partial{\cal B}={\cal M}$, and define the WZ action as an integral over the ball by 
\begin{equation*}
S_{\rm wz}=\int_{\cal B}\alpha=\int_{\cal M}\beta \ .
\end{equation*}
There are actually inequivalent possibilities for the ball over which the WZ action is defined, say ${\cal B}$  and ${\cal B}'$.  In order for the physics to be insensitive to the choice made, the difference between the two actions must be a integer multiple of $2\pi$,  
\begin{align}
     \int_{\cal B}\alpha-\int_{{\cal B}'}\alpha&=\int_{S^{d+1}}\alpha=2\pi k\, ,\  k\in\mathbb{Z}\ ,\label{WZquantizationcondition}
\end{align}
so that the path integral is unaffected.     As indicated in \eqref{WZquantizationcondition}, the difference between the integrals over the two balls is equivalent to a single integral over the $(d+1)$-sphere generated by gluing the two balls together.  Essentially, one has compactified $\mathcal{M}$ into the $d$-sphere, $S^{d}$, which is the equator of a $(d+1)$-sphere, $S^{d+1}$, whose northern hemisphere is  $\mathcal{B}$, and whose southern hemisphere is $\mathcal{B}'$.  The $(d+1)$-form $\alpha$ now defines a map from $S^{d+1}$ into the coset space $G/H$. Such maps are classified by the homotopy group $\pi_{d+1}(G/H)$; if this group is nontrivial, then it is possible for $\int_{S^{d+1}}\alpha\neq 0$ and the condition~\eqref{WZquantizationcondition} forces the coefficient of $\alpha$ to be quantized.  The coupling constant therefore cannot change continuously and hence it cannot be renormalized.    This is the procedure of \cite{Witten:1983tw} where it is shown that the Wess--Zumino--Witten term of the chiral lagrangian must be built in this manner and enjoys a non-renormalization theorem.  See \cite{D'Hoker:1994ti} for more details and subtleties in such constructions.
 
\subsection{Spacetime Symmetry Breaking}

The coset treatment for spontaneously broken spacetime symmetries proceeds much as the internal case, but with two main subtleties.  For simplicity, we take $\mathcal{M}$ to be $d$-dimensional Minkowski space in this section.

The first subtlety is that every translation generator is included in the coset element $\tilde{g}$, regardless of whether it's truly broken or not.  The reason for this is a practical one; the generators included in the coset element are precisely those that induce non-linear symmetry transformations, and since translations act non-linearly on coordinates {\it i.e.}, $x^{\mu}\mapsto x^{\mu}+b^{\mu}$, they too need to appear in the coset element, despite the fact that they may act linearly on fields in the theory.

The coset elements are then members of $G/H$, where $H$ includes all preserved transformations \textit{except} for translations.  Generators of $H$ and preserved translations are denoted by $V_{I}$ and $P_{\mu}$, respectively, and the remaining generators are denoted by $Z_{a}$.  We write the coset element as
\be
\tilde{g}=\exp(x^{\mu}P_{\mu})\exp(\xi^{a}Z_{a})~,
\ee 
where $x^{\mu}$'s are the spacetime coordinates.  The Maurer--Cartan 1-form is expanded as 
\be
\Omega=\Omega_{Z}+\Omega_{V}+\Omega_{P}=\Omega_{Z}^{a}Z_{a}+\Omega_{V}^{I}V_{I}+\Omega_{P}^{\mu}P_{\mu}~,
\ee
and $\Omega_{V}$ again transforms as a connection, while $\Omega_{P}$ and $\Omega_{Z}$ transform homogeneously.  Further, $\Omega_{P}$ defines a vielbein for the system with components $(\Omega_{P})_\mu^{\nu}$ defined by $\Omega_{P}=\rd x^{\mu}(\Omega_{P})_\mu^{\nu}P_{\nu}$, where $\nu$ is the Lorentz index, so that the covariant metric is given by $g_{\mu\nu}=(\Omega_{P})_\mu^{\alpha}(\Omega_{P})_\nu^{\beta}\eta_{\alpha\beta}$.

The second subtlety is that, in the case of spacetime symmetry breaking, there can be fewer Goldstone modes than broken symmetries.  There are various interpretations of this phenomenon, see  \cite{Ivanov:1975zq,Nielsen:1975hm,Low:2001bw,McArthur:2010zm,Watanabe:2011ec,Hidaka:2012ym,Watanabe:2013iia,Nicolis:2013sga,Endlich:2013vfa,Brauner:2014aha}, but there is no general consensus and it is a topic of ongoing research.  
In any case, a procedure exists for determining when one can reduce the number of degrees of freedom by eliminating fields in the action.  Schematically, the rule is that if the commutator of an unbroken translation generator, $P$, and a broken generator, $Z_{1}$, contains another broken generator, $Z_{2}$, that is $[P,Z_{1}]\sim Z_{2}$, then it is possible to eliminate the field corresponding to $Z_{1}$ in favor of the remaining fields and their derivatives. The relation between fields is determined by setting parts of the Maurer--Cartan form along $Z_{2}$ to zero. This is known as the {\it Inverse Higgs} (IH) effect \cite{Ivanov:1975zq}. In practice the elimination of the $Z_{1}$ field is often equivalent to integrating the field out via its equations of motion~\cite{McArthur:2010zm}, but this is not always the case \cite{Goon:2012dy}.  

 There is something of an art to choosing which parts of the Maurer--Cartan form to set to zero and in determining whether the fields ought to be eliminated at all \cite{Nicolis:2013sga}, but the only requirement from a consistency standpoint is that the final theory obey all of the symmetries contained within $G$. We take the viewpoint that one chooses which inverse Higgs constraints to apply based on the degrees of freedom one wishes to describe.  For example, when considering Yang--Mills we know that we are interested in gauge bosons and therefore we eliminate all higher order fields in favor of $A_{\mu}^{a}$.  Keeping the other fields may be interesting in other contexts, but not the one we wish to study here, and we leave such explorations to future work.

\section{Yang--Mills As A Nonlinear Realization}

With the knowledge of the previous sections, we can construct the YM action on $d$-dimensional Minkowski space.  Much of the following section is a modern rephrasing of the original calculation of \cite{Ivanov:1976pg} and appears elsewhere in the literature (see~\cite{Riccioni:2009hi} for a nice review), but some lesser known results will be emphasized. Although the results are very general, we will assume that the gauge group is a simple group for simplicity.
For these groups, one can choose a matrix representation for the generators, $\{T_{a}\}$, which satisfy ${\rm tr}(T_{a}T_{b})\propto \delta_{ab} = {\rm diag}(+,\ldots,+)$.\footnote{In addition, for a simple group, the structure constants of the Lie algebra can be made totally anti-symmetric by employing the Cartan--Killing metric $g_{ab} = f_{ad}^{~~c}f_{bc}^{~~d}$.} We shall use the notations interchangably when convenient. The existence of other invariant tensors depends on the group in question and for the majority of our purposes it is sufficiently general to only contract group indices with $\delta_{ab}$.

\subsection{The Local Yang--Mills Algebra}

We now construct the algebra which the gauge fields nonlinearly realize. Although our main interest is non-Abelian gauge theory, it is helpful for building intuition to first consider the Abelian case. An Abelian 1-form gauge field, $A$, transforms under a gauge transformation, with $g = e^{\alpha(x) Q}$ as
\be
A \longmapsto g^{-1}(A+\rd)g = A + \rd \alpha~,
\ee
where $Q$ is the U(1) generator and $\alpha(x)$ is an arbitrary function. We can imagine Taylor expanding this function as
\be
\alpha(x) Q = \sum_{n=0}^\infty c_{\mu_1\cdots\mu_n}x^{\mu_1}\cdots x^{\mu_n}Q~,
\ee
where the $c_{\mu_1\cdots\mu_n}$ are constant coefficients. If we then define new generators
\be
Q^{\mu_1\cdots\mu_n} \equiv x^{\mu_1}\cdots x^{\mu_n}Q~,
\ee
we can think of a gauge transformation as being built out of an infinite number of global rotations~\cite{Riccioni:2009hi}
\be
g = e^{ c Q+\sum_{n=1}^\infty c_{\mu_1\cdots\mu_n}Q^{\mu_1\cdots\mu_n}}~.
\ee
The interpretation is then that the global transformation generated by $Q$ is linearly realized, while the transformations generated by the $Q^{\mu_1\cdots\mu_n}$ are nonlinearly realized and the gauge field is the corresponding Goldstone boson. Notice that the generators $Q^{\mu_1\cdots\mu_n}$ explicitly depend on $x^\mu$ and therefore do {\it not} commute with the spacetime Poincar\'e generators.

With this intuition, we can now proceed to construct the algebra nonlinearly realized by a non-Abelian gauge field. We denote the generators of the global part of the algebra by $T_{a}$, $a\in\{1,\ldots,N\}$, satisfying commutation relations 
\be
[T_{a},T_{b}]=-gf_{ab}^{~~c}T_{c}~,
\ee
where $g$ is the gauge coupling. Latin gauge indices are raised and lowered with $\delta_{ab}$ and whether an index is up or down is unimportant.  The Poincar\'e generators are $P_{\mu}$ and $J_{\mu\nu}$.  
 
In order to define local gauge generators we again expand the gauge generator as $\alpha^{a}(x)T_{a}\equiv \sum_{n=0}^{\infty}\alpha^a_{\nu_1\ldots\nu_n}x^{\nu_{1}}\ldots x^{\nu_n}T_{a}$ for some set of constants $\alpha_{\nu_{1}\ldots\nu_{n}}^a$.  Then, defining $T^{\nu_{1}\ldots\nu_{n}}_a\equiv x^{\nu_{1}}\ldots x^{\nu_{n}}T_{a}$, these generators obey commutation relations\footnote{We symmetrize with weight one, {\it e.g.}, $T_{(\mu\nu)}=\frac{1}{2}\left (T_{\mu\nu}+T_{\nu\mu}\right )$}
\begin{align}
[T^{\alpha_{1}\ldots\alpha_{n}}_a,T^{\beta_{1}\ldots\beta_{m}}_b]&=-g f_{ab}^{~~c}T^{\alpha_{1}\ldots\alpha_{n}\beta_{1}\ldots\beta_{m}}_c\nn
[P_{\mu},T^{\nu_{1}\ldots\nu_{n}}_a]&=-n\delta_{\mu}^{(\nu_1}T^{\nu_{2}\ldots\nu_{n})}_a\ ,\label{localGalgebra}
\end{align}and indeed we take \eqref{localGalgebra} to define the algebra under study, along with the relations for $[P_{\mu},J_{\alpha\beta}]$, $[J_{\mu\nu},J_{\rho\sigma}]$ and $[J_{\mu\nu},T^{\alpha_{1}\ldots\alpha_{n}}_a]$ whose specific forms will not be needed.  We call $\{T^{\alpha_{1}\ldots\alpha_{n}}_a\}$, $n\neq 0$, the ``local generators" of the gauge group and $\{T_{a}\}$ are the ``global generators." The union of the two sets generates $G_{\rm local}$ while $\{T_{a}\}$ generates $G_{\rm global}\subset	G_{\rm local}$.

\subsection{Unbroken Phase}

When the global symmetry is preserved, the $SO(3,1)$ rotations and $G_{\rm global}$ transformations act linearly on gauge fields and YM is reproduced by studying the ``breaking" pattern
\be
G_{\rm local}\times ISO(3,1)
\longrightarrow G_{\rm global}\times SO(3,1)~.
\ee  
The representative coset element belongs to $(G_{\rm local}\times ISO(3,1))/(G_{\rm global}\times SO(3,1))$ and is written as
\begin{align}
\!\tilde{g}\!=\!e^{x^{\mu}P_{\mu}}\!\left [\ldots\right ]\!e^{\Phi_{\alpha_{1}\alpha_{2}\alpha_{3} }^aT^{\alpha_{1}\alpha_{2}\alpha_{3} }_a} \! e^{\Phi_{\nu_{1}\nu_{2}}^bT^{\nu_{1}\nu_{2}}_b}\! e^{-A_{\mu}^cT_c^{\mu }} \ , 
\label{unbrokenrepresentativecosetelement}
\end{align}
where the fields $\{\Phi_{\nu_1\ldots\nu_n }^a\}$ are totally symmetric in all Greek indices and the terms in $[\ldots]$ are all higher order in that they contain generators with more Greek indices.  

\subsubsection{Maurer--Cartan Form and Inverse Higgs}

Given the coset representative~\eqref{unbrokenrepresentativecosetelement}, we can compute the Maurer--Cartan form and expand it as $\Omega=\Omega_{P}^{\mu}P_{\mu}+\Omega^{a}T_{a}+\Omega^a_{\nu }T_a^{\nu}+\ldots$, with components calculated to be 
\begin{align}
\Omega_{P}^{\mu}&=\rd x^{\mu}\nn
\Omega^{a}&= \rd x^{\nu}A_{\nu}^a\nn
\Omega_{\nu}^a&=-\rd A_{\nu}^a-2\rd x^{\mu}\Phi_{\mu\nu}^a+\frac{1}{2}gf_{bc}^{~~a}A_{\mu}^bA_{\nu }^c\rd x^{\mu}~.\label{UnbrokenYMMCform}
\end{align}
The commutation relation $[P_{\mu},T^{\alpha\beta}_a]=-2\delta_{\mu}^{(\alpha}T^{\beta)}_a$ reveals that we can eliminate the field corresponding to $T^{\alpha\beta}_a$, {\it i.e.}, $\Phi_{\alpha\beta}^a $, through an inverse Higgs constraint. In components,\footnote{There is potential confusion here as Greek indices are used both as form indices and indices on the fields, but their meaning should be clear in context.} we have $\Omega_{\nu}^a=\rd x^{\mu}\Omega_{\mu\nu}^a$, which we can separate into a symmetric and an anti-symmetric piece, where $\Phi_{\alpha\beta}^a$ only appears in the symmetric components
\begin{align}
\Omega_{(\mu\nu)}^a&=-\partial_{(\mu}A_{\nu)}^a-2\Phi_{\mu\nu }^a\nn
\Omega_{[\mu\nu]}^a&=-\partial_{[\mu}A_{\nu]}^a+\frac{1}{2}gf^{~~a}_{bc}A_{\mu }^bA_{\nu }^c=-\frac{1}{2}F_{\mu\nu }^a\ .\label{(anti)symmetricpartsofMCform}
\end{align}
Setting $\Omega_{(\mu\nu)}^a=0$, we eliminate $\Phi_{\mu\nu }^a$ in favor of derivatives of $A_{\mu }^a$ through
\be
\Phi_{\mu\nu }^a=-\frac{1}{2}\partial_{(\mu}A_{\nu)}^a~.
\ee
Evaluating $\Omega_{\nu }^a$ on this constraint, we obtain the YM field strength tensor, denoted
\be
\Omega_{\nu }^a\big\rvert_{\rm IH}=-\frac{1}{2}F_{\mu\nu}^a\rd x^{\mu}~.
\ee

A similar pattern holds when performing the calculation to higher orders.  The fields $\Phi_{\alpha_{1}\ldots\alpha_{n}}^a$ can all be removed by IH constraints which eliminate them in favor of $A_{\nu}^a$ and its derivatives.  The higher order components of the Maurer--Cartan form all turn into gauge covariant derivatives of the field strength tensor~\cite{Riccioni:2009hi}.

\subsubsection{Symmetries}

As a consistency check, we confirm that actions constructed using the IH constraint still respect the full set of $G_{\rm local }$ symmetries.  We perform a generic symmetry transformation by acting on $\tilde{g}$ with $\exp(-\sum_n\epsilon_{\alpha_1\ldots\alpha_{n} }^aT^{\alpha_{1}\ldots\alpha_{n}}_a)$.  Defining $\epsilon^{a}(x)\equiv \sum_n x^{\nu_{1}}\ldots x^{\nu_{n}}\epsilon_{\nu_{1}\ldots\nu_{n}}^a$, the infinitesimal transformations are found to be
\begin{align}
\delta A_{\mu }^a&=gf_{bc}^{~~a}\epsilon^{b}A_{\mu }^c+\partial_{\mu}\epsilon^{a}\nn
\delta \Phi_{\mu\nu }^a&= gf_{bc}^{~~a}\epsilon^{b}\Phi_{\mu\nu }^c-\frac{1}{2}\partial_{\mu}\partial_{\nu}\epsilon^{a}-\frac{1}{2}gf_{bc}^{~~a}A^c_{(\nu}\partial_{\mu)}\epsilon^{b} \ .\label{gaugeparametersymmetry}
\end{align}
The first line of \eqref{gaugeparametersymmetry} is recognized as the familiar non-Abelian gauge transformation with gauge parameter $\epsilon^{a}$. The second line demonstrates that $-\frac{1}{2}\partial_{(\mu}A_{\nu)}^a$ and $\Phi_{\mu\nu }^a$ have the same transformation properties and hence the IH replacement, $\Phi^a_{\mu\nu }\mapsto-\frac{1}{2}\partial_{(\mu}A_{\nu)}^a$, is consistent and yields an action invariant under the full $G_{\rm local}$ group.

\subsubsection{Construction of the Action}

We now construct the action. As discussed previously, the components of the Maurer--Cartan form along the preserved generators define the connection used in matter covariant derivatives, and thus from \eqref{UnbrokenYMMCform} the connection is simply $A^a_{\mu}\rd x^{\mu}\equiv A^{a}$, as expected.  The remaining components are contracted in $H$-invariant ways.  For the case under consideration, this simply translates into the requirement that Latin gauge indices are contracted with factors of $\delta_{ab}$ and Greek indices are contracted with $\eta_{\mu\nu}$ or $\epsilon_{\mu_1\ldots\mu_{d}}$.

We are imposing the IH constraint $\Phi_{\mu\nu }^a=-\frac{1}{2}\partial_{(\mu}A_{\nu)}^a$ and so the lowest order forms we can build actions with are $\Omega_{P}^{\mu}=\rd x^{\mu}$ and $\Omega_{\nu }^a\big\rvert_{\rm IH}=-\frac{1}{2}F^a_{\mu\nu }\rd x^{\mu}$.   In generic $d>2$ dimensions, there is only one possible invariant action which is quadratic in $\Omega_{\nu}^a\big\rvert_{\rm IH}$,
\begin{align}
\mathcal{L}_{2}&\propto  \delta_{ab}F^{a}\wedge\star F^{b} =  {\rm tr}~F\wedge\star F~,
\label{quadraticYMlagrangianunbrokenphase}
\end{align} where $F$ is given by\footnote{This relation is actually independent of whether or not one chooses to impose the IH constraint.}
\be 
F^{a}=\frac{1}{2}F_{\mu\nu }^a\rd x^{\mu}\wedge\rd x^{\nu}= \Omega_{P}^{\nu}\wedge\Omega_{\nu }^a
\ee 
and where $\star$ is the Hodge star with respect to the vielbein on $\mathcal{M}$ defined by $\Omega_{P}^{\mu}$ (here just the flat metric).  This is nothing but the standard Yang--Mills kinetic term, ${\rm tr}~F\wedge\star F \propto \rd^{d}x\, F_{\mu\nu }^aF^{\mu\nu}_a$, and hence the coset construction is seen to generate the correct term.

We therefore see that the objects from which we can construct an invariant action are the field strength tensor $F_{\mu\nu}^aT_a$ and the gauge covariant derivative $D_\mu = \partial_\mu +ig A_\mu^a T_a $, and that any Lagrangian ${\cal L}(D_\mu , F_{\mu\nu})$ built from these ingredients will be gauge invariant, which is a familiar result.

In even dimensions, there are further possibilities. For example, in $d=4$ there can be a term of the form $\mathcal{L}_{\theta}\propto {\rm tr}~F \wedge F$.  This is nothing but the topological $\theta$-term which is known to be a total derivative.  It is straightforward to verify this fact by explicitly writing out $\mathcal{L}_{\theta}$ in terms of $A_{\mu}^{a}$, but it proves more interesting and fruitful to  verify this fact through an alternative procedure which takes advantage of the underlying group structure and makes connections to standard techniques of coset methods.  
\subsubsection{Chern--Simons as a Wess--Zumino Term }
Given a group $G$ with Lie algebra generators $X_{a}$ which obey $[X_{a},X_{b}]=f_{ab}^{~~c}X_{c}$, the components of the Maurer--Cartan form obey the Maurer--Cartan equation,
\begin{align}
\rd\Omega^{c}+\frac{1}{2}f_{ab}^{~~c}\Omega^{a}\wedge\Omega^{b}=0\ ,
\end{align}
where $\Omega=\Omega^{a}X_{a}$.
For the Lie algebra at hand \eqref{localGalgebra} and the Maurer--Cartan components in \eqref{UnbrokenYMMCform}, this implies
\begin{align}
\rd\Omega_{P}^{\mu}&=0\nn
\rd \Omega^{a}&=\frac{g}{2}f_{bc}^{~~a}\Omega^{b}\wedge\Omega^{c}+\Omega_{P}^{\mu}\wedge\Omega_{\mu}^a\nn
\rd\Omega_{\mu}^a&=gf_{bc}^{~~a}\Omega_{\mu}^{ b}\wedge\Omega^{c}+\Omega_{P}^{\nu}\wedge\Omega_{\nu\mu}^a\ .\label{exteriorderivativerelationsunbrokencase}
\end{align}
The exterior derivative relations \eqref{exteriorderivativerelationsunbrokencase} then provide the necessary ingredients for the cohomological problem of finding Wess--Zumino terms.  

For simplicity, we will focus on the case $d=4$ and use \eqref{exteriorderivativerelationsunbrokencase} to show that ${\rm tr}~F\wedge F$ is closed and, relatedly, that there exists a WZ term in $d=3$ which is simply the Chern--Simons term. We follow the general WZ strategy and look for closed 4-forms. There is no need to impose IH for this part of the calculation; we consider the form
\be
\mathcal{L}_{\theta}\equiv\delta_{ab}(\Omega_{P}^{\mu}\wedge\Omega_{\mu}^a)\wedge(\Omega_{P}^{\nu}\wedge\Omega_{\nu}^b)~.
\ee
Using the relations~\eqref{exteriorderivativerelationsunbrokencase} we find that this form is closed,
\begin{align}
\rd\mathcal{L}_{\theta} = 2gf_{ca}^{~~d}(\Omega_{P}^{\mu}\wedge\Omega_{\mu}^c)\wedge(\Omega_{P}^{\nu}\wedge\Omega_{\nu }^a)\wedge\Omega_{d}=0\ ,
\end{align}
since under $c\leftrightarrow a$ the structure constant is antisymmetric while the forms are symmetric.  As indicated by our notation, this 4-form is simply the $\theta$-term from the previous section, $\mathcal{L}_{\theta}\propto {\rm tr} ~F\wedge F$, and thus we have proven the claim that it is a total derivative using the coset framework. 

Further, $\mathcal{L}_{\theta}$ is the exterior derivative of a $3$-form which is not itself $H$-invariant.  Explicitly, 
 \begin{align}
 \mathcal{L}_{\theta}&=\rd\left [\Omega^{\mu}_{P}\wedge\Omega_{\mu}^{a}\wedge\Omega^{b}\delta_{ab}+\frac{1}{6}gf_{abc}\Omega^{a}\wedge\Omega^{b}\wedge\Omega^{c}\right ]\equiv \rd\mathcal{L}_{{\rm cs}_3} \ ,\label{CSintermsofOmegas}
 \end{align}
and since we are not allowed to use $\Omega^{a}$ in constructing the relevant $H$-invariant actions, $\mathcal{L}_{{\rm cs}_3}$ is not itself strictly $H$-invariant.  Therefore, $\mathcal{L}_{{\rm cs}_3}$ represents a WZ term which can appear in the $d=3$ action which would have been missed in a purely three dimensional coset construction.  Replacing the $\Omega$'s in favor of $A^{a}$ and the field strength, we find
 \begin{align}
 \mathcal{L}_{{\rm cs}_3}&=\delta_{ab}A^{a}\wedge F^{b}+\frac{1}{6}gf_{abc}A^{a}\wedge A^{b}\wedge A^{c}\nn
 &=\delta_{ab}A^{a}\wedge\rd A^{b}-\frac{1}{3}gf_{abc} A^{a}\wedge A^{b}\wedge A^{c} \ ,
 \label{ChernSimons3form}
 \end{align}
as the notation indicates, and as one may have expected, the Wess--Zumino 3-form for non-Abelian gauge fields is simply the Chern--Simons term.
  Note that this result is in harmony with the earlier discussion about symmetries in the coset construction; the Chern--Simons term (a WZ term) changes by an exact form under a gauge transformation while the standard, non-WZ, kinetic term is {\it strictly} invariant.\footnote{It is well known that the coupling of the $d=3$ non-Abelian Chern--Simons form is quantized for certain gauge groups \cite{Dunne:1998qy}. In particular, the $d=3$ Chern--Simons level is quantized if the homotopy group $\pi_3(G)$ is nontrivial.  On the other hand, a naive application of Wess-Zumino-Witten \cite{Witten:1983tw} type arguments would alternatively indicate that this coupling is quantized if $\pi_4\left ((G_{\rm local}\times ISO(2,1))/(G_{\rm global}\times SO(2,1)\right )$ is non-trivial.  Further explorations of the relation between the two arguments would be interesting, but fall outside the scope of this paper.} A similar connection between Chern--Simons and Wess--Zumino terms was noted in~\cite{Brauner:2014ata}.

 The extension to higher dimensional cases is straightforward, but group dependent.  If the gauge group $G_{\rm global}$ admits an invariant tensor of the form $M_{a_{1}\ldots a_{k}}$, then there exists a closed, gauge-invariant $2k$-form $\mathcal{L}^{2k}$ which will lead to WZ terms defined by
  \begin{align}
 \mathcal{L}^{2k}=\rd\beta^{2k-1}_{\rm wz}=M_{a_{1}\ldots a_{k}}\Omega^{\mu}_{P}\wedge\Omega_{\mu}^{a_1}\wedge\ldots\wedge\Omega^{\nu}_{P}\wedge \Omega_{\nu}^{a_k}\ ,\label{genericCSforms}
 \end{align} but the existence of such an invariant tensor depends on the precise gauge group at hand.  In terms of the usual treatment of gauge fields, given the set of generators $\{T_{a}\}$ the tensor $M_{a_{1}\ldots a_k}$ exists if ${\rm tr}(T_{a_1}\ldots T_{a_k})$ is non-vanishing.  Assuming the tensor exists, then the $(2k-1)$-form $\beta^{2k-1}_{\rm wz}$ shifts under the gauge transformation by an exact form.  Having chosen the gauge group and determined the values of $k$ for which \eqref{genericCSforms} exists, we can build the WZ terms appropriate for odd, $d=2n-1$ dimensions.  
 
 There are, in general, multiple such terms and they are all constructed by wedging a single $\beta^{2k-1}_{\rm wz}$ type form with a set of $\mathcal{L}^{2k}$ type forms such that the final result is a $(2n-1)$-form.  That is,
 \be 
{\cal L}_{\rm wz} = \beta^{2k-1}_{\rm wz}\wedge\mathcal{L}^{2k_{2}}\wedge\ldots\wedge\mathcal{L}^{2k_m}
 \ee
will define a WZ term (equivalently, a Chern--Simons term) in $d=2n-1$, so long as $k_1+\ldots+k_m=n$ and the appropriate invariant tensors exist to generate each of these forms.  For instance, defining the shorthand $\Omega_{(2)}^{a}\equiv \Omega_{P}^{\mu }\wedge\Omega_{\mu }^{a}$, $d=7$ will always inherit a Chern--Simons term from the 8-form
 \begin{align}
 \mathcal{L}^{8}{}'&\equiv \delta_{ab}\delta_{cd}\Omega_{(2)}^{a}\wedge\Omega_{(2)}^{b}\wedge\Omega_{(2)}^{c}\wedge\Omega_{(2)}^{d} \ ,\label{CS8form1}
 \end{align}
 and if $M_{abcd}$ exists then there will also be a second Chern--Simons term coming from
 \begin{align}
 \mathcal{L}^{8}&\equiv M_{abcd}\Omega_{(2)}^{a}\wedge\Omega_{(2)}^{b}\wedge\Omega_{(2)}^{c}\wedge\Omega_{(2)}^{d}\ .\label{CS8form2}
 \end{align}
 
These results are well-known, so this section is intended both as a translation between the language and concepts of coset constructions and those of gauge theories and as a further check that the methods of non-linear realizations can reproduce known results.   Indeed the final two lines of \eqref{exteriorderivativerelationsunbrokencase} are nothing but the definition of the curvature tensor and the Bianchi identity, respectively implying
\begin{align}
&F^{a} = \rd A^{a}-\frac{1}{2}gf_{bc}^{~~a}A^{b}\wedge A^{c}\nn
&\rd F^{a} - gf_{bc}^{~~a} A^{b}\wedge F^{c}=0\ .\label{explicitbianchiidentity}
\end{align}
If one were to repeat the calculations of this section using the identities in the form \eqref{explicitbianchiidentity} rather than \eqref{exteriorderivativerelationsunbrokencase}, the derivation of Chern--Simons terms would have directly mirrored methods familiar in the literature.  For instance, the 8-forms in \eqref{CS8form1} and \eqref{CS8form2} simply correspond to
\begin{align}
\mathcal{L}_{8}{}'&\propto \left ({\rm tr} F\wedge F\right )\wedge\left ({\rm tr} F\wedge F\right )\nn
\mathcal{L}_{8}&\propto {\rm tr}(F\wedge F\wedge F\wedge F)~,
\end{align}
in the usual language, with corresponding Chern--Simons 7-forms.

\subsection{Spontaneously Broken Phase}

We now want to address the situation in which the gauge symmetry is spontaneously broken. For concreteness, we consider the case where \textit{all} of the gauge symmetries are broken. It is straightforward to apply the following procedure to more general breaking patterns and we will comment on other scenarios in a later section.

The primary difference from the unbroken case is that now the global part of the gauge transformation is {\it also} nonlinearly realized. Therefore, only Lorentz symmetry is linearly realized, and the corresponding breaking pattern is $G_{\rm local}\times ISO(3,1)\to SO(3,1)$. The representative coset element belongs to $(G_{\rm local}\times ISO(3,1))/SO(3,1)$ and can be written as
\begin{align}
\tilde{g}&=e^{x^{\mu}P_{\mu}}\left [\ldots\right ]e^{\Phi_{\nu_1\nu_2}^bT^{\nu_{1}\nu_{2}}_b}e^{-A_{\mu}^cT^{\mu}_c}e^{\pi^{a}T_{a}}\label{brokenrepresentativecosetelement}\ ,
\end{align}
in the same manner as \eqref{unbrokenrepresentativecosetelement}.

\subsubsection{Maurer--Cartan Form and Inverse Higgs}

The construction of the Maurer--Cartan form and the implementation of the inverse Higgs constraint proceed along the same lines as the unbroken case.  Since the representative coset element of the broken case is related simply to that of the unbroken case, {\it i.e.}, $\tilde{g}_{\rm broken}=\tilde{g}_{\rm unbroken}e^{\pi^{a}T_{a}}$, the two Maurer--Cartan forms are also closely related
\begin{align}
\Omega_{\rm broken}\! =\!e^{-\pi^{a}T_{a}}\Omega_{\rm unbroken}e^{\pi^{a}T_{a}}+e^{-\pi^{a}T_{a}}\rd e^{\pi^{a}T_{a}}\ .\label{MCbrokenintermsofMCunbroken}
\end{align} 
Expanding out the Maurer--Cartan form as before, we find that the coefficients are
\begin{align}
\Omega_{P}^{\mu}&=\rd x^{\mu}\nn
\Omega^{a}&= \rd x^{\nu}A_{\nu}^bU(\pi)_{b}^a+\frac{1}{g}f_{bc}^{~~a}U(\pi)_{d}^b\rd U(\pi)^{dc}\nn
\Omega_{\nu}^a&=-\rd A_{\nu}^bU(\pi)_{b}^a-2\rd x^{\mu}\Phi_{\mu\nu}^bU(\pi)_{b}^a+\frac{1}{2}gf_{bc}^{~~d}A_{\mu}^bA_{\nu}^c\rd x^{\mu}U(\pi)_{d}^a\ ,\label{MCcomponentsbrokenphase}
\end{align}
where we have defined the matrix
\begin{align}
U(\pi)_{a}^b&=\delta_{a}^b+g\pi^{c}f_{ca}^{~~b}+\frac{1}{2}g^{2}\pi^{c}\pi^{c'}f_{ca}^{~~d}f_{c'd}^{~~~b}+\ldots=\exp[g\pi^{c}f_{ca}^{~~b}]~,\label{DefinitionU(pi)}
\end{align}
also note that $U^{-1}(\pi)_{a}^b=U(\pi)_{b}^a$. Additionally, we have chosen to normalize the generators so that $f_{ac}^{~~d}f_{bd}^{~~c}=\delta_{ab}$.

The IH constraint which eliminates $\Phi_{\alpha\beta }^a$ remains the same:\footnote{Notice that the commutation relation $[P_{\mu},T^{\nu}_a]=-\delta_{\mu}^{\nu}T_{a}$ implies that it is possible to eliminate $A_{\mu}^a$ in favor of $\pi^{a}$ and its derivatives through an inverse Higgs constraint. In accordance with our philosophy on the IH effect, we \textit{choose} not to implement the constraint.  Physically, we know that we want to describe gauge bosons so we keep the $A_{\mu}^{a}$'s.  In practice, the resulting IH constraint would force $A_{\mu}^{a}$ to be pure gauge, resulting in a trivial Maurer--Cartan form free of any dynamical fields, which is indeed invariant under the relevant symmetries, but is not particularly useful.} $\Phi_{\alpha\beta}^a=-\frac{1}{2}\partial_{(\alpha}A_{\beta)}^a$.  After imposing this constraint, we have $\Omega_{\nu}^a\big\rvert_{\rm IH}=-\frac{1}{2}\rd x^{\mu}F_{\mu\nu}^bU(\pi)_{b}^a$.

\subsubsection{Relation to the St\"uckelberg Trick}

The above calculation demonstrates that the ingredients derived from coset methods for building Yang--Mills actions in the Higgs phase are simply a realization of the St\"uckelberg trick used to restore gauge symmetries. In this section, we make the correspondence explicit.

First, we review the implementation of the St\"uckelberg trick in the theory of massive $SU(N)$ YM gauge bosons which, for simplicity, all have the same mass $m$,
\begin{align}
\mathcal{L}&=-\frac{1}{4g^{2}}{\rm tr}F_{\mu\nu}F^{\mu\nu}-\frac{m^{2}}{2g^{2}}{\rm tr} A_{\mu}A^{\mu}\ .\label{MassiveYMforStuckelbergSection}
\end{align}

This lagrangian is not gauge invariant, but gauge invariance can be restored by coupling in new fields, $\pi^{a}(x)$, with $a\in\{1,\ldots,N^2-1\}$, {\it i.e.}, one field for each generator of $SU(N)$.  In order to insert the $\pi^{a}(x)$'s appropriately, one first performs a gauge transformation with $\pi^{a}(x)$ as the gauge parameter,
\begin{align}
A_{\mu}&\longmapsto U^\dagger(\pi)(A_{\mu}+\partial_{\mu})U(\pi)\equiv A'_{\mu}\nn
F_{\mu\nu}&\longmapsto U^\dagger(\pi) F_{\mu\nu}U(\pi)\equiv F'_{\mu\nu}\label{StuckelbergTransformationExample}
\end{align}
where $U(\pi)=e^{\pi^{a}(x)T_{a}}$ is an element of $SU(N)$.  We then define a new lagrangian $\mathcal{L}'$ by taking \eqref{MassiveYMforStuckelbergSection} and replacing $A_{\mu}\mapsto A_{\mu}'$ and $F_{\mu\nu}\mapsto F_{\mu\nu}'$.  That is,
\begin{align}
\mathcal{L}'&=-\frac{1}{4g^{2}}{\rm tr}F'_{\mu\nu}F'^{\mu\nu}-\frac{m^{2}}{2g^{2}}{\rm tr}A'_{\mu}A'^{\nu}\nn
&=\!-\frac{1}{4g^{2}}\!{\rm tr}F_{\mu\nu}F^{\mu\nu}\!-\!\frac{m^{2}}{2g^{2}} {\rm tr}D_{\mu}U(\pi)D^{\mu}U^\dagger(\pi)~, \label{StuckelbergedLagrangianEx}
\end{align}
where $D_{\mu}U(\pi)= \partial_{\mu}U(\pi)+A_{\mu}U(\pi)$ is the gauge covariant derivative of $U(\pi)$.  The lagrangian $\mathcal{L}'$ then enjoys a gauge symmetry under which we simultaneously change
\begin{align}
A_{\mu}&\longmapsto V^\dagger(x)(A_{\mu}+\mathds{1}\partial_{\mu})V(x)\nn
U(\pi)&\longmapsto V^\dagger(x)U(\pi)~,
\end{align}
where $V(x)\in SU(N)$.

The physics of the $\mathcal{L}$ and $\mathcal{L}'$ lagrangians is the same, we have just made the degrees of freedom in~\eqref{MassiveYMforStuckelbergSection} manifest.  In $\mathcal{L}'$, we introduced $N^{2}-1$ new fields, but also restored $N^{2}-1$ gauge symmetries and hence degree of freedom counting is the same for both cases. We can demonstrate the equivalence explicitly by using the gauge symmetry of $\mathcal{L}'$ to go ``unitary gauge" in which we set $U(\pi)\to \mathds{1}$, where the two lagrangians coincide.  

The above process and its generalizations are known collectively as the St\"uckelberg trick, which is often a useful tool for elucidating the physics in certain regimes of theories, especially at high energies. See \cite{Burgess:1992gx,Hinterbichler:2011tt,Preskill:1990fr,ArkaniHamed:2002sp} for good discussions of St\"uckelberg fields in various contexts.

The St\"uckelberged fields in \eqref{StuckelbergTransformationExample} are precisely the terms which arise in the Maurer--Cartan form \eqref{MCcomponentsbrokenphase} when applied to completely broken $SU(N)$.  We noted in \eqref{MCbrokenintermsofMCunbroken} that the Maurer--Cartan forms for the broken and unbroken phases of YM are related in a very simple manner.
If we expand out the terms on each side of~ \eqref{MCbrokenintermsofMCunbroken} using
\begin{align}
\Omega_{\rm broken}&=\Omega'{}_{P}^{\mu }P_{\mu}+\Omega'{}^{a}T_{a}+\Omega'{}_{\nu}^aT^{\nu}_a+\ldots\nn
\Omega_{\rm unbroken}&=\Omega_{P}^{\mu }P_{\mu}+\Omega^{a}T_{a}+\Omega_{\nu}^aT^{\nu}_a+\ldots
\end{align} 
then \eqref{MCbrokenintermsofMCunbroken} demonstrates that
\begin{align}
\Omega'{}^{a}T_{a}&=e^{-\pi^{b}(x)T_{b}}\big(\Omega^{a}T_{a}+\rd\big )e^{\pi^{b}(x)T_{b}}\nn
\Omega'{}_{\nu}^aT^{\nu}_a&=e^{-\pi^{b}(x)T_{b}} \Omega_{\nu}^aT^{\nu}_ae^{\pi^{b}(x)T_{b}}\ .
\end{align}
Identifying $\Omega_{a}\mapsto A_{\mu}^a\rd x^{\mu}$, $\Omega_{\nu}^a\mapsto -\frac{1}{2}F_{\mu\nu}^a\rd x^{\nu}$ and $e^{\pi^{b}(x)T_{b}}=U(\pi)$ (and similar for primed terms), we easily see that the ingredients we obtain from the Maurer--Cartan form \eqref{MCcomponentsbrokenphase} are precisely the same as the fields used in the St\"uckelberg trick \eqref{StuckelbergTransformationExample}.

  The St\"uckelberg trick is applicable to more general theories (in our example the gauge masses were only chosen to be equal for simplicity) and, as we will see, the broken phase Maurer--Cartan components \eqref{MCcomponentsbrokenphase} provide the completely general building blocks for generating the low energy action in completely broken gauge theories.

It is interesting that the coset construction for broken YM theories automatically comes replete with St\"uckelberg fields, but perhaps it is not surprising.  After all, the coset construction generates actions which non-linearly realize every broken symmetry of the system.  Massive gauge theories written in the standard form, as in \eqref{MassiveYMforStuckelbergSection}, retain none of the broken gauge symmetries.  Only when we couple in St\"uckelberg fields do we restore a realization of gauge invariance and hence one might have expected this method to coincide with the coset result.

\subsubsection{Relation to Gauging Cosets by Hand}

Finally, we comment on the relation between our construction, the St\"uckelberg trick and the method in which coset models are gauged by hand \cite{Burgess:1992gx,Weinberg:1996kr}.  This technique explores the role of gauge fields in SSB systems essentially by using the St\"uckelberg trick in reverse.  Rather than beginning with the gauge fields, one starts with the Goldstones $\{\pi^{a}\}$ which parameterize the coset space $G/H$ corresponding to a global symmetry breaking pattern $G\to H$.  

As detailed previously, the action for the bosons is generated by writing a typical element of $G/H$ as $\tilde g=e^{\pi^{a}T_{a}}$, taking the $\pi^{a}$'s to be the Goldstone fields and defining the non-linearly realized symmetries of $\pi^{a}$ by $\pi^{a}\to \pi'^{a}$ via the relation
\begin{align}
 ge^{\pi^{a}T_{a}}&=e^{\pi'^{a}T_{a}}h(\pi,g)\ ,\label{cosettransformationHandsection}
 \end{align}
 or equivalently $g:\tilde{g}\to \tilde{g}'=g\tilde{g}h^{-1}(\pi,g)$,
 where $g$ is an arbitrary, spacetime independent element of $G$.  One then constructs the Maurer--Cartan form $\Omega=\tilde g^{-1}\rd \tilde g=\Omega_{Z}+\Omega_V=\Omega^{a}_{Z}Z_{a}+\Omega^{I}_{V}V_{I}$, where $V_{I}$'s generate $H$ and $Z_a$ are the remaining, broken generators. Under \eqref{cosettransformationHandsection} these components transform as
 \begin{align}
   g:\begin{cases} \Omega_{Z}&\longmapsto h(\pi,g)\Omega_{Z}h^{-1}(\pi,g)\\
   \Omega_{V}&\longmapsto h(\pi,g)\left (\Omega_{V}+\rd\right )h^{-1}(\pi,g)\end{cases}\ .\label{byhandsectionMCtransformations}
   \end{align}
   
   If we wish to promote this to local transformations, $g\to g(x)$, so that the transformation now replaces $\pi^{a}$ by $\pi^{a}\to \pi'^{a}$ via the new relation
   \begin{align}
 g(x)e^{\pi^{a}T_{a}}&=e^{\pi'^{a}T_{a}}h(\pi,g(x))\ ,\label{cosettransformationHandsectionlocal}
 \end{align}
 or equivalently $g(x):\tilde{g}\mapsto\tilde{g}'=g(x)\tilde{g}h^{-1}(\pi,g(x))$,
 we need to introduce a gauge field $A$ with components along all of the generators of $G$.  Since we wish to insert $A$ in such a way that we retain the nice properties we had when working with the Maurer--Cartan form, it proves useful to consider the object $\tilde{\Omega}\equiv \tilde{g}^{-1}(\rd +A)\tilde{g}=\Omega+\tilde{g}^{-1}A\tilde{g}$.  Demanding that under the action of $g(x)$, $A$ transforms as $g(x):A\mapsto g(x)(A+\rd)g(x)^{-1}$, we find that the total transformation of $\tilde{\Omega}$ is
 \begin{align}
 g(x):\tilde{g}^{-1}(\rd +A)\tilde{g}&\mapsto h \tilde{g} ^{-1}g(x)^{-1}\left [\rd +g(x)Ag(x)^{-1}+g(x)\rd g(x)^{-1}\right ]g(x)\tilde{g}h^{-1}\nn
 &=h\tilde\Omega h^{-1}+h\rd h^{-1}\ .
 \end{align} 
 If we break up $\tilde{\Omega}$ into its components along $Z_a$ and $V_{I}$ as $\tilde{\Omega}=\tilde \Omega_{Z}+\tilde \Omega_{V}=\tilde \Omega_{Z}^{a}Z_{a}+\tilde \Omega_{V}^{I}V_{I}$ then, entirely analogously to the ungauged case \eqref{byhandsectionMCtransformations}, the above transformation law and the assumed properties of the groups imply that the components of $\tilde{\Omega}$ along broken generators transform homogeneously while those along unbroken generators transform as a connection,
 \begin{align}
 g(x):\begin{cases}\tilde{\Omega}_{Z}&\longmapsto h(\pi,g(x))\tilde{\Omega}_{Z}h(\pi,g(x))^{-1}\\
 \tilde{\Omega}_{V}&\longmapsto h(\pi,g(x))(\tilde{\Omega}_{V}+\rd )h(\pi,g(x))^{-1}\end{cases}\ .
 \end{align}
 From here, the coset construction proceeds as normal; one uses $\tilde{\Omega}_{Z}$ to write down $H$-invariant actions, and in particular one can use these terms to write down masses for the broken gauge generators.  In the limiting case where the gauge group is entirely broken, we see that we are required to build actions with $\tilde{g}^{-1}(\rd+A)\tilde{g}$, with $\tilde{g}\in G$, which is exactly the object we used to build actions in the previous section when using the St\"uckelberg trick which was, in turn, found to be essentially equivalent to the construction used in this paper.
 
  In the end, the building blocks found through the methods of this section, the St\"uckelberg trick and the spacetime coset techniques presented in this paper are identical to each other, but rather than starting with only gauge fields or only Goldstone bosons, our method incorporates both simultaneously and fits entirely within the framework of non-linear realizations, which may prove to be a technical advantage.  In particular, when studying Higgs mechanisms for systems which also spontaneously break Poincar\'e invariance, spacetime coset methods handle the implementation of inverse Higgs constraints quite naturally and would seem to be better suited for these scenarios than St\"uckelberging or gauging by hand would be~\cite{GHJTGravitypaper}.

\subsubsection{Construction of the Action}

We now construct the $d$-dimensional action appropriate for the broken phase of the theory. Two crucial differences between the broken and unbroken cases are that we can now use $\Omega^{a}$ in the construction of the action (because it now corresponds to a broken generator) and we no longer are required to contract gauge indices with $\delta_{ab}$.

Defining $\Omega_{P}^{\nu}\wedge\Omega_{\nu}^a\equiv\mathcal{F}^{a}=  F^{b}\Omega(\pi)_{b}^a$, the most general $SO(3,1)$ invariant action quadratic in the Maurer--Cartan components is
\begin{align}
\mathcal{L}_{2}&= I_{ab}\mathcal F^{a}\wedge\star \mathcal F^{b}-M_{ab}\Omega^{a}\wedge\star\Omega^{b}+\Theta_{ab}\mathcal F^{a}\wedge \mathcal F^{b}+\varepsilon_{ab}\Omega^{a}\wedge\Omega^{b}\label{GeneralBrokenYMLagrangian}
\end{align}
for arbitrary tensors $I_{ab}$ and $M_{ab}$.  The tensors $\Theta_{ab}$ and $\varepsilon_{ab}$ can be non-zero only in $d=4$ and $d=2$, respectively, and we do not consider them further. Any theory whose gauge group is completely broken will have a low energy description whose action takes on the form \eqref{GeneralBrokenYMLagrangian}, at lowest order.

Basic physical requirements place coarse constraints on the form of \eqref{GeneralBrokenYMLagrangian}, but finer grained information cannot be determined without further assumptions.  The requirement that there be no tachyons forces the mass matrix $M_{ab}$ to be positive semi-definite, while freedom from ghosts and technical naturalness\footnote{Here we mean requiring that gauge invariance be restored in the limit that the gauge masses are taken to zero.  This ensures that corrections to the gauge masses are proportional to the masses themselves, so that the mass is not raised to the cutoff by quantum corrections.} require that  $I_{ab}$  parametrically reduces to $I_{ab}\to \lambda\delta_{ab}$, with $\lambda>0$, as $M_{ab}\to 0$~\cite{Hinterbichler:2011tt}. But there is no physical principle that determines, for example, the distribution of the gauge boson masses.  
	
This is as it should be, since the knowledge that the symmetry group is entirely broken is not enough information on its own to determine the distribution of gauge masses. The coset construction only gives us invariant objects that can be used to build low-energy effective actions and finding a UV theory which gives rise to specific parameters is a separate question. In fact, even given a micophysical model where the field content, interactions and couplings are specified, this information may not be enough to uniquely determine the spectrum of gauge masses. For instance, there could be moduli in the theory which break the gauge symmetries by acquiring vacuum expectation values (VEVs), which in turn set the gauge masses.  Since there exist many possible values for the VEVs, there exist many possible distributions for the gauge masses.   

The coset construction can thus properly reproduce the generic form of the low energy action for a completely spontaneously broken gauge theory.  We found that the gross features of the action could be determined, but certain details could not.  This is in part an artifact of the example we chose to study. Often the gauge group is not entirely broken, rather it is only broken down to a subgroup, or there may be additional UV global symmetries which  more tightly constrain the form of the low energy action than in the setting considered above.  We briefly touch on some of these alternative situations in the next section.

\subsubsection{WZ Terms in the Broken Phase}

For completeness, we search for possible Wess--Zumino terms in the broken phase.  Unlike in the unbroken case, there are no terms which we a priori expect to find.  That is, the Chern--Simons term fit all of the criteria for a WZ term and it was fairly clear that it would arise as such in the unbroken phase. However, there are no well known analogues which appear in the action only in the broken phase and, indeed, we will not find any WZ terms here.

The procedure is nearly identical to that of the unbroken case.  For simplicity we assume that in addition to $G_{\rm local}$ there is an additional global $G$ symmetry and that the total group is spontaneously broken down to the diagonal subgroup\footnote{Along with the usual ``breaking" pattern for the Poincar\'e symmetries, which we have omitted writing here.}, $G_{\rm local}\times G\to (G_{\rm local}\times G)_{\rm diag}$.  This is not a significant change.  The Maurer--Cartan components for this pattern are the same as in \eqref{MCcomponentsbrokenphase} and the preserved symmetry just forces us to contract Latin gauge indices with gauge invariant tensors, as we will see more explicitly in later examples.  The choice is simply made for brevity, since it reduces the number of terms we need to consider when constructing actions.

We start with the components of the Maurer--Cartan form and use the Maurer--Cartan structure equations to find $H$-invariant, closed $(d+1)$-forms which are locally the exterior derivative of a $d$-form which is not itself $H$-invariant.  The $d$-form defines a $d$-dimensional WZ term.  Since the algebra under study remains unchanged in the broken phase, all of the Maurer--Cartan structure equations remain exactly the same.  The only differences are that the Maurer--Cartan components now have different dependences on the fields\footnote{For example, in the broken phase $\Omega^{a}=\rd x^{\nu}A_{\nu}^bU(\pi)_{b}^a+\frac{1}{g}f_{bc}^{~~a}U(\pi)_{d}^b\rd U(\pi)^{dc}$, while in the unbroken phase it was simply $\Omega^{a}=\rd x^{\mu}A_{\mu}^{a}$.} and we can now use $\Omega^{a}$, whereas it was forbidden in the unbroken case.  Finally, to be precise, we concentrate on the construction of WZ terms in $d=3$ in order to compare to the work of previous sections.

We find that in the broken phase there exist only two invariant $4$-forms that are generically closed\footnote{By ``generically closed", we mean closed for all possible gauge groups.  Certain forms may end up being closed only for particular gauge groups simply due to dimensionalities. For example, $\Omega^{a}\wedge\Omega^{b}\wedge\Omega^{c}\wedge\Omega^{d}$ vanishes for $SU(2)$ because there are only three independent gauge indices, and this fact could in principle cause certain forms to be closed when studying $SU(2)$, but this would not generally be true.}. The first such $4$-form is
\begin{align}
 \mathcal{L}&=\Omega_{\mu}^{a}\wedge\Omega^{\mu}_{P}\wedge\Omega^{b}\wedge\Omega^{c}f_{abc}\ ,
 \end{align} 
 but this is the exterior derivative of an allowed $3$-form,
 \begin{align}
 \mathcal{L}&=\frac{1}{3}\rd\left [f_{abc}\Omega^{a}\wedge\Omega^{b}\wedge\Omega^{c}\right ] \ ,
 \end{align}
 The second $4$-form is familiar to us: it is simply the $\theta$-term
 \begin{align}
\mathcal{L}_{\theta}&\equiv \delta_{ab}\left (\Omega_{P}^{\mu}\wedge\Omega_{\mu}^{a}\right )\wedge\left (\Omega_{P}^{\nu}\wedge\Omega_{\nu}^{b}\right )\ ,
\end{align}
and as we stated earlier, we already knew this would be closed in the broken phase since the Algebra and form of the Maurer-Cartan structure equations remain unchanged.  Again, we previously found \eqref{CSintermsofOmegas} that $\mathcal{L}_{\theta}$ is exact
 \begin{align}
\mathcal{L}_{\theta}&=\rd\left [\Omega^{\mu}_{P}\wedge\Omega_{\mu}^{a}\wedge\Omega^{b}\delta_{ab}+\frac{1}{6}g\Omega^{a}\wedge\Omega^{b}\wedge\Omega^{c}f_{abc}\right ]= \rd\mathcal{L}_{{\rm cs}_3}\ ,\label{Broken3DCSterm}
\end{align}
but now the conclusion is different.  In the unbroken phase the use of $\Omega^{a}$ was disallowed in the coset construction and so $\mathcal{L}_{3}$ was missed when attempting to generate $3$-form actions, meaning that $\mathcal{L}_{{\rm cs}_3}$ represented a true WZ term.  In the broken phase, $\Omega^{a}$ can now be used to generate $3$-forms and so $\mathcal{L}_{{\rm cs}_3}$ is simply a $3$-form action that can be written down within the usual coset framework.  In the language of non-linear realizations, $\mathcal{L}_{{\rm cs}_3}$ no longer represents a WZ term in the broken phase.  Therefore, there are no WZ terms at all for spontaneously broken $d=3$ gauge theories.

There is some potential confusion with respect to the interpretation of $\mathcal{L}_{{\rm cs}_3}$ in the Higgs phase which deserves comment.  We no longer have $\Omega^{a}T_{a}=A^{a}T_{a}$, but instead, from \eqref{MCcomponentsbrokenphase}, it is of the form $\Omega^{a}T_{a}=U(\pi)^{-1}\left (A^{a}T_{a}+\rd\right )U(\pi)$, where $U(\pi)$ is an element of $G$ which depends on the $\pi^{a}$ fields.  Therefore, $\mathcal{L}_{{\rm cs}_3}$ defined in \eqref{Broken3DCSterm} is the normal Chern--Simons 3-form \eqref{ChernSimons3form} with the replacement $A\mapsto U^{-1}(A+\rd)U$ everywhere, which we write as $\mathcal{L}_{{\rm cs}_3}[U^{-1}\left (A+\rd\right )U]$.  Since the CS term is gauge invariant up to a total derivative it is tempting to replace $\mathcal{L}_{{\rm cs}_3}[U^{-1}\left (A+\rd\right )U]\to \mathcal{L}_{{\rm cs}_{3}}[A]$ and remove the St\"uckelberg fields entirely.  Possible confusion arises as this replacement obscures the fact that the CS term is no longer a Wess-Zumino term in the sense that it should no longer shift by a total derivative.  That is, we have claimed that non-WZ terms are strictly invariant, but if we make the above replacement then we will find that $\mathcal{L}_{{\rm cs}_3}[A]$ shifts by a total derivative under gauge transformations, as usual.  The key is that to preserve the non-WZ nature of this term in the broken phase we should not drop the total derivatives we get when removing the St\"uckelberg fields as they are responsible for keeping $\mathcal{L}_{{\rm cs}_3}[U^{-1}\left (A+\rd\right )U]$ strictly gauge invariant.  For most purposes, either form of $\mathcal{L}_{{\rm cs}_3}$ will be fine and this discussion is just a clarification on the internal consistency of these coset procedures.  However, there does exist one subtlety in that the usual non-renormalization argument \cite{Dunne:1998qy} for the CS coupling constant does not go through for $\mathcal{L}_{{\rm cs}_3}[U^{-1}\left (A+\rd\right )U]$ as it crucially relies on the fact that the CS term shifts by a total derivative.

\section{Applications and Generalizations}

In this section, we generalize the previous construction in two ways. First, we consider a situation where the gauge symmetry is not completely broken, but rather is broken to some subgroup. Additionally, we consider the generalization of our techniques to $p$-form gauge theories, and construct actions for these theories in both the unbroken and St\"uckelberg phases using coset techniques.

\subsection{Other Breaking Patterns}
First we consider symmetry breaking patterns where the UV physics contains both gauged and global copies of a symmetry group $G$ which is spontaneously broken down to a group which contains a diagonal version of $G$. 

As a concrete example, consider an $SU(N)$ theory with $N$ Higgs fields transforming in the fundamental representation.  We can combine the Higgs fields into an $N\times N$ matrix $\Phi$ and build an appropriate potential out of ${\rm tr}[\Phi^{\dagger}\Phi]$ such that the fields acquire a vacuum expectation value $\langle\Phi\rangle\propto \mathds{1}$~\cite{Preskill:1990fr}.  The potential has an $SU(N)\times SU(N)\times U(1)$ symmetry under which $\Phi\to e^{i\theta}L\Phi R$, where $e^{i\theta}\in U(1)$ and $L,R\in SU(N)$.
 
Now imagine that we have gauged the left transformations.  The VEVs then generate the breaking pattern $SU(N)_{\rm local}\times SU(N)_{\rm global}\times U(1)_{\rm global}\to SU(N)_{\rm diagonal}$, since $\langle\Phi\rangle\to L\langle\Phi\rangle R$ with $L=R^{-1}$ is the only preserved symmetry.
 
 We now examine this breaking pattern using our coset methods.  Letting the generators of $SU(N)_{\rm local}$, $SU(N)_{\rm global}$ and $U(1)_{\rm global}$ be $\{T^{\nu_1\ldots\nu_n}_a\}$, $\{U_{a}\}$ and $V$ respectively, the only preserved internal symmetry is generated by the diagonal set $\{T_{a}-U_{a}\}$.  It is convenient to take the basis of broken generators as $\{T^{\nu_{1}\ldots\nu_n}_a,V\}$, in which case the representative coset element is
 \begin{align}
 \tilde{g}'{}&=e^{x^{\mu}P_{\mu}}[\ldots]e^{\Phi_{\mu\nu}^aT^{\mu\nu}_a}e^{-A_{\mu}^cT^{\mu}_c}e^{\pi^{a}T_{a}}e^{\phi V}\ ,
 \end{align}
 and we may expand the Maurer--Cartan form as
 \begin{align}
 \tilde{g}'{}^{-1}\rd \tilde{ g}'{}&=\Omega'{}_{P}^{\mu}P_{\mu}+\Omega_T'{}^{a}T_{a}+\Omega_U'{}^{a}U_{a}+\Omega'_VV+\Omega_T'{}_{\nu}^aT^{\nu}_a+\ldots\ .
 \end{align}
The calculation is almost the same as in the broken case \eqref{MCcomponentsbrokenphase}, and the result is
 \begin{align}
 \Omega_P'{}^{\mu}&= \rd x^{\mu}\nn
 \Omega_V'&=\rd\phi \nn
 \Omega'_T{}^{a}&=A_{\nu}^b\rd x^{\nu}U(\pi)_{b}^a+\frac{1}{g}f_{bc}^{~~a}U(\pi)_{d}^b\rd U(\pi)^{dc}\nn
 \Omega'_T{}_{\nu}^a&=-\rd A_{\nu}^bU(\pi)_{b}^a-2\rd x^{\mu}\Phi_{\mu\nu}^bU(\pi)_{b}^a +\frac{1}{2}gf_{bc}^{~~d}A_{\mu}^bA_{\nu}^c\rd x^{\mu}U(\pi)_{d}^a \ , 
 \label{brokendiagonalexampleMCform}
 \end{align}
 where the matrix $U(\pi)_{a}^b$ is defined in \eqref{DefinitionU(pi)}.

Every part of the Maurer--Cartan form lies along a broken generator, since there is no component along $U_{a}$, and hence we can use every component of \eqref{brokendiagonalexampleMCform} to build actions.  It is crucially important that there is now a preserved global, diagonal $SU(N)$ symmetry, as this dictates that the Latin gauge indices must be contracted with $\delta_{ab}$.  This eliminates many of the $U(\pi)_a^b$ factors, since $U(\pi)_{a}^bU(\pi)_{b}^c = \delta_a^c$, and the most general, stable action that is quadratic in the components of the Maurer--Cartan form is found to be
\begin{align}
\mathcal{L}&=-\frac{1}{4g^{2}}{\rm tr} F_{\mu\nu}F^{\mu\nu}-\frac{m^{2}}{2g^{2}}{\rm tr}D_{\mu}U(\pi)D^{\mu}U^{-1}(\pi)-\frac{1}{2}(\partial\phi)^{2}\ ,
\end{align}
where we have imposed the inverse Higgs constraint, employed trace notation rather than displaying the explicit $\delta_{ab}$'s and have used the gauge covariant derivative notation of \eqref{StuckelbergedLagrangianEx}.  

Therefore, the result of the symmetry breaking pattern is $N^{2}-1$ gauge bosons of equal mass and a massless Goldstone field corresponding to the broken $U(1)$, in accord with the results of \cite{Preskill:1990fr}.\footnote{We have assumed that the acquired gauge boson masses are smaller
than the masses of the radial modes so that the massless $\phi$ field
and the gauge fields are the lowest energy degrees of freedom.}  We thus see that the coset methods applied to gauge theories can have more predictive power when more symmetries are preserved.   Also note that had we not included the $U(1)$ factor we would have reproduced the Lagrangian of \eqref{StuckelbergTransformationExample}.

\subsection{Generalization to $p$-forms}

Thus far, we have been studying one-form gauge fields, but it is also straightforward to generalize to the case of Abelian $p$-forms.  Such a form, $A_p = A_{\mu_1\cdots\mu_{p}}\rd x^{\mu_1}\wedge\cdots\wedge\rd x^{\mu_p}$, transforms under a gauge transformation as
\be
A_p\longmapsto A_p+\rd \Lambda_{p-1}~,
\ee
where $\Lambda_{p-1}$ is a $(p-1)$-form. Explicitly, in components, this is
$A_{\mu_1\cdots\mu_p} \mapsto A_{\mu_1\cdots\mu_p}+\partial_{[\mu_1}\Lambda_{\mu_2\cdots\mu_p]}$.
As before, the global part of the gauge transformation (for which $\rd\Lambda = 0$) is linearly realized on $A_p$, while the local part of the gauge transformation is realized non-linearly.
Similar to the Yang--Mills case, we can expand out the gauge parameter $\Lambda_{p-1}$ and define new generators by
$T^{\alpha_1\cdots\alpha_n \mu_1\cdots \mu_{p-1}} \equiv x^{\alpha_1}\cdot\cdots x^{\alpha_n} T^{\mu_1\cdots\mu_{p-1}}$,
all of which commute with each other. Note that $T^{\mu_1\cdots\mu_{p-1}}$ generates the global part of the gauge transformation, while the other generators generate the local transformations. These generators have non-trivial commutators with spacetime translations
\be
\left[ P_\nu, T^{\alpha_1\cdots\alpha_n \mu_1\cdots \mu_{p-1}}\right] = -n\delta_\nu^{(\alpha_1}T^{\alpha_2\cdots\alpha_n) \mu_1\cdots \mu_{p-1}}~,
\ee
and also with spacetime rotations and boosts, but the form of this latter commutator will be immaterial to our purposes.

We are interested in the coset $G_{\rm local}/G_{\rm global}$, which we parameterize as
\be
\tilde g = e^{x^\mu P_\mu}\cdots e^{\Phi_{\alpha_1\alpha_2\mu_1\cdots\mu_{p-1}}T^{\alpha_1\alpha_2\mu_1\cdots\mu_{p-1}}}e^{A_{\alpha\mu_1\cdots\mu_{p-1}}T^{\alpha\mu_1\cdots\mu_{p-1}}}~.
\ee
Note that $A_{\alpha\mu_1\cdots\mu_{p-1}}$ is anti-symmetric in all of its indices.
From this, we can compute the components of the Maurer--Cartan form
\be
\Omega = \Omega^\mu P_\mu + \Omega_{\mu_1\cdots\mu_{p-1}}T^{\mu_1\cdots\mu_{p-1}}+\Omega_{\alpha\mu_1\cdots\mu_{p-1}}T^{\alpha\mu_1\cdots\mu_{p-1}}
\ee
where the coefficients are given by
\begin{align}
\Omega^\mu &= \rd x^\mu\\
\Omega_{\mu_1\cdots\mu_{p-1}} &= \rd x^\alpha A_{\alpha\mu_1\cdots\mu_{p-1}}\\
\Omega_{\alpha\mu_1\cdots\mu_{p-1}} &= \rd A_{\alpha\mu_1\cdots\mu_{p-1}} -2\rd x^\beta\Phi_{(\alpha\beta)\mu_1\cdots\mu_{p-1}}  \ .
\end{align}
As before, we can eliminate the field $\Phi$ through an inverse Higgs constraint by setting
\be
\frac{1}{2}\partial_{(\beta}A_{\alpha)\mu_1\cdots\mu_{p-1}} -2\rd x^\beta\Phi_{(\alpha\beta)\mu_1\cdots\mu_{p-1}} = 0~.
\label{pformIH}
\ee
This projects the part symmetric in $(\alpha\beta)$. Upon substituting back in, we obtain the field strength:\footnote{Here we have used the fact that $\partial_{[\beta} A_{\alpha]\mu_1\cdots\mu_{p-1}} = \partial_{[\beta} A_{\alpha\mu_1\cdots\mu_{p-1}]}$, because $A$ is antisymmetric in all its indices.}
\be
\Omega_{\alpha\mu_1\cdots\mu_{p-1}} = \frac{1}{2}\rd x^\beta\partial_{[\beta} A_{\alpha\mu_1\cdots\mu_{p-1}]} = \frac{1}{2} \rd x^\beta F_{\beta\alpha\mu_1\cdots\mu_{p-1}}~.
\ee
Using this, we can construct the quadratic action
\be
{\cal L}_2 = \Omega_{\mu_1\cdots\mu_{p}} \wedge \star\Omega^{\mu_1\cdots\mu_{p}} \sim  F_{p+1}\wedge\star F_{p+1} \ ,
\ee
where $F_{p+1} = \rd A_p$.
Notice that in the case $p=1$, with $A = A_\mu\rd x^\mu$, this reduces precisely to Maxwell electrodynamics, as expected. 

Similar to the Yang--Mills case, there exist Wess--Zumino terms for $p$-forms, descending from topological terms built out of products of $\Omega^{\mu_1\cdots\mu_{p}}$, of the form
\be
{\cal L} \sim F_{p+1}\wedge F_{p+1}\wedge\cdots\wedge F_{p+1}~,
\ee
which exist whenever $n(p+1) = d$ for some $n$. These terms are exact (${\cal L} = \rd\beta_{\rm cs}$), and have corresponding Wess--Zumino terms in one lower dimension, which are again the Chern--Simons terms
\be
\beta_{\rm cs} = A_{p}\wedge F_{p+1}\wedge\cdots\wedge F_{p+1}~,
\ee
which shift by a total derivative under a gauge transformation.

\subsubsection{Spontaneously broken $p$-form gauge theories}

We now consider the theory of spontaneously broken $p$-form gauge fields. Much as in the Yang--Mills case, in order to construct the theory in the broken phase, we consider breaking the global part of the field transformation also, corresponding to $T^{\mu_1\cdots\mu_{p-1}}$, which has a corresponding Goldstone boson, $B_{\mu_1\cdots\mu_{p-1}}$. We therefore consider the coset element
\be
\tilde g = \tilde g_{\rm local}e^{B_{\mu_1\cdots\mu_{p-1}}T^{\mu_1\cdots\mu_{p-1}}}~.
\ee
The Maurer--Cartan forms are similar to before: after imposing the inverse Higgs constraint~\eqref{pformIH}, the MC 1-forms are
\begin{align}
\Omega^\mu &= \rd x^\mu\\
\Omega_{\mu_1\cdots\mu_{p-1}} &= \rd x^\alpha\left(A_{\alpha\mu_1\cdots\mu_{p-1}}+\partial_{[\alpha}B_{\mu_1\cdots\mu_{p-1}]}\right)\\
\Omega_{\alpha\mu_1\cdots\mu_{p-1}} &= \frac{1}{2} \rd x^\beta F_{\beta\alpha\mu_1\cdots\mu_{p-1}}~.
\end{align}
Notice that in addition to the kinetic term $\sim F^2$, we can also now construct a mass term for the gauge field, so that the quadratic lagrangian is (in form notation)
\be
{\cal L} = -\frac{1}{2}\rd A_p\wedge\star \rd A_p-\frac{m^2}{2}(A_p+\rd B_{p-1})\wedge\star(A_p+\rd B_{p-1})~,
\ee
which is invariant under a gauge transformation where both $A_p$ and $B_{p-1}$ transform
\begin{align}
A_p&\longmapsto A_p+\rd \Lambda_{p-1}~,\\
B_{p-1} &\longmapsto B_{p-1} - \Lambda_{p-1}~.
\end{align}
Here the form field $B_{p-1}$ is a St\"uckelberg field which restores gauge invariance in the massive theory. This type of St\"uckelberg realization of the gauge symmetry arises, for example, in the worldvolume action of D-branes in the presence of background $p$-form gauge fields~\cite{Polchinski:1998rq}.
Again, in the case $p=1$, this reduces to the scalar St\"uckelberg field that appears in a massive $U(1)$ theory.

\subsection{Anomalous Gauge Theories}

Finally, we note that the discussion has so far remained classical, but the above methods could have applications to interesting quantum mechanical aspects of symmetry breaking.  There exist gauge theories which are anomaly free in the UV, but which display gauge anomalies in the IR after SSB occurs, as is nicely discussed in \cite{Preskill:1990fr}.  Here, a gauge invariant description can be restored by coupling in St\"uckelberg fields through a Wess--Zumino term whose gauge variation does not vanish, but precisely cancels the anomalous variation arising from the other low energy fields.  It would be an interesting exercise to explore whether the physics of such anomalous low energy theories can be captured in the coset language.

\section{Conclusion}

The coset methods of Callan, Coleman, Wess and Zumino, and Volkov \cite{Callan:1969sn,Coleman:1969sm, volkov} have proven to be invaluable tools for exploring the low energy behavior of systems which exhibit the spontaneous breaking of internal symmetries. These methods were later extended and it was shown that the YM lagrangian naturally arises if one studies a ``breaking" pattern in which a local gauge symmetry is ``broken" to the global group \cite{Ivanov:1976pg,Borisov:1974bn}.  In this paper, we have extended these methods to study the case where gauge symmetries are truly spontaneously broken.

First, we reproduced the coset construction of Yang--Mills gauge theories in modern language.  The results are familiar, but it is interesting to approach YM from this non-standard direction.  For instance, the search for a Wess--Zumino action in the coset calculation, in the sense of \cite{D'Hoker:1994ti}, was shown to lead to the Chern--Simons terms.  

Next, we have extended these techniques to the case where even the global symmetry is non-linearly realized, which physically corresponds to a true breaking of the gauge symmetry.  Coset methods faithfully reproduce the result that the low energy physics is described by massive gauge bosons.  Depending on the breaking pattern, it can be possible to discern the distribution of gauge boson masses while remaining agnostic about the UV physics.  Historically this is what makes the coset construction powerful: generic properties of the low energy physics can be discerned from only the knowledge of the breaking pattern.   A search for WZ terms in the broken phase revealed that none exist.

We were able to reproduce familiar results about St\"uckelberged Yang--Mills theories, where the non-linear realization of gauge symmetry arises in an interesting way. Further, we were able to reproduce results about other breaking patterns, where only some of the gauge symmetries are broken in this alternative language. Similar analyses can be performed for other, related patterns.  For instance, the above procedure could be used to study color flavor locking~\cite{Alford:1998mk}, electroweak symmetry breaking via chiral condensates, or theories with approximate custodial symmetries.  Alternatively, it is straightforward to study examples in which subsets of the gauge group are preserved. Additionally, it was straightforward to generalize our results to the case of $p$-form gauge theories. Although we focused on Abelian $p$-form theories, it would be interesting to see if these techniques could be applied to the construction of actions for non-Abelian $p$-form gauge fields.

Many other directions in which to generalize present themselves. An obvious one is to consider what happens if in addition to gauge symmetry, spacetime symmetries are broken. In this note we have assumed that Poincar\'e symmetry is preserved by the symmetry breaking physics, but this need not be so, and the formalism should generalize readily to this case. Such Higgs phases of non-relativistic gauge theories have recently been considered in~\cite{Watanabe:2014qla,Gongyo:2014sra}, and it would be intetresting to try to reproduce their results in this language. Another possible application of these techniques is to the duality recently pointed out in~\cite{deRham:2014lqa} for Abelian vector fields. Similar to the way that the duality enjoyed by galileon theories~\cite{deRham:2013hsa} can be understood from the coset perspective~\cite{Creminelli:2014zxa, Kampf:2014rka}, it should be possible to understand this vector duality using the techniques presented here. Concretely, the duality should follow upon the identification $T_\mu \mapsto T_\mu+\alpha P_\mu$, which implements the field redefinition $x^\mu \mapsto x^\mu+ \alpha A^\mu$ in the low-energy theory.

A final application of these methods concerns studying Higgs phases of gravity. Various proposals for Higgs mechanisms of gravity have been previously suggested, but only with the recent discovery of dRGT \cite{deRham:2010kj} has it become known how to construct an apparently consistent, ghost free theory of massive gravity, making it the leading candidate for describing a potential Higgs phase of gravity.  Applying our methods to general relativity, it is possible to determine to what extent dRGT can be expected as the generic low energy description of spontaneously broken gravity. These avenues are currently under investigation~\cite{GHJTGravitypaper}.

\noindent
{\bf Acknowledgments}:
We thank Lasha Berezhiani, Kurt Hinterbichler, Denis Klevers, Emil Martinec, Alberto Nicolis, Riccardo Penco, Rachel Rosen and Yi-Zen Chu for useful dicussions. The work of G.G. and M.T. was supported in part by the US Department of Energy. The work of A.J. was supported in part by the Kavli Institute for Cosmological Physics at the University of Chicago through grant NSF PHY-1125897, an endowment from the Kavli Foundation and its founder Fred Kavli, and by the Robert R. McCormick Postdoctoral Fellowship.

\bibliographystyle{utphys}
\bibliography{BrokenYangMillsCoset}

\end{document}